\documentclass[fp]{jpsj3}
\usepackage{txfonts}
\usepackage{color}

\title{Investigation of Transport Properties for FeSe$_{1-x}$Te$_x$ Thin Films under Magnetic Fields}

\author{Yuichi Sawada, Fuyuki Nabeshima, Yoshinori Imai\thanks{imai@tohoku.ac.jp}\footnote{Present address: Department of Physics, Tohoku University, 6-3, Aramaki Aza-Aoba, Aoba-ku, Sendai 980-8578, Japan}, and Atsutaka Maeda}
\inst{Department of Basic Science, the University of Tokyo, 3-8-1 Komaba, Meguro-ku, Tokyo 153-8902, Japan} 

\abst{We investigated the transport properties under magnetic fields of up to 9 T for FeSe$_{1-x}$Te$_x$ thin films on CaF$_2$. Measurements of the temperature dependence of the electrical resistivity revealed that for $x = 0.2 - 0.4$, where $T_{\rm c}$ is the highest, the width of the superconducting transition increased with increasing magnetic field, while the width was almost the same with increasing magnetic field for $x = 0 - 0.1$. In addition, the temperature dependence of the Hall coefficient drastically changed between $x = 0.1$ and $0.2$ at low temperatures. These results indicate that clear differences in the nature of the superconductivity and electronic structure exist between $x=0-0.1$ and $x \ge 0.2$.}

\begin{document}
\maketitle

The discovery of iron-based superconductors has triggered much attention for fundamental studies and applications \cite{Kamihara}. One of the iron-based superconductors, FeSe has the simplest crystal structure, composed of conducting planes alone\cite{Hsu}. Although the superconducting transition temperature, $T_{\rm c}$, of FeSe is 8 K, which is low compared with other iron-based superconductors, $T_{\rm c}$ reaches as high as 30 K under high pressure \cite{Masaki,Medvedev}. In addition, monolayer FeSe films on SrTiO$_3$ substrates exhibit very high $T_{\rm c}$ \cite{Wang}. 
Although it is under debate whether these high $T_{\rm c}$ have the same origin or not, these results demonstrate that FeSe has potential as a high-$T_{\rm c}$ superconductor. To raise its $T_{\rm c}$, the partial substitution of Te for Se is also effective. FeSe$_{1-x}$Te$_x$ has $T_{\rm c}$ of up to 14 K at $x = 0.5 - 0.6$ \cite{Fang}. In addition, the fabrication of thin films makes $T_{\rm c}$ for FeSe$_{0.5}$Te$_{0.5}$ higher than that of bulk crystals \cite{Bellingeri,Iida,TsukadaAPEX}. Therefore, the fabrication of FeSe$_{1-x}$Te$_x$ thin films is important for both fundamental studies and applications.

It is known that single-phase bulk samples with $0.1 \leq x \leq 0.4$ cannot be obtained because of a phase separation, and this fact has prevented the complete understanding of FeSe$_{1-x}$Te$_x$ \cite{Fang}. Recently, we have succeeded in obtaining FeSe$_{1-x}$Te$_x$ thin films with these compositions on CaF$_2$ substrates by pulsed laser deposition (PLD) \cite{ImaiPNAS}. $T_{\rm c}$ for these films increases with decreasing $x$ for $0.2 \leq x \leq 1$ and reaches 23 K at $x = 0.2$. This value is 1.5 times higher than the highest value obtained for bulk samples. Surprisingly, we observed the sudden suppression of $T_{\rm c}$ between $x = 0.1$ and $0.2$. Therefore, it is of great interest to investigate the difference in the physical properties other than $T_{\rm c}$ in these ranges of $x$. 

In this letter, we will show the temperature dependence of the electrical resistivity under magnetic fields and the Hall effect for FeSe$_{1-x}$Te$_x$ thin films in order to clarify how the transport properties change as a function of $x$. Our results show that the superconducting transition width, upper critical field, and Hall coefficient change in the range $x = 0.1 - 0.2$, suggesting that a definite change indeed takes place in the electronic structure of FeSe$_{1-x}$Te$_x$ in this range of $x$.

\begin{table}
\caption{Specifications of the FeSe$_{1-x}$Te$_x$ thin films.}
\label{tab:SampleSpec}
\begin{center}
\begin{tabular}{cccccc}
\hline\hline
 $x$&Thickness (nm)&$a$ (\AA)&$c$ (\AA)&$T_{\rm c}^{\rm onset}$ (K)&$T_{\rm c}^{\rm zero}$ (K) \\
\hline
0&197&3.735&5.584&14.6&13.2 \\ 
0.1&77&3.753&5.585&11.5&10.1 \\
0.2&41&3.749&5.710&22.6&20.3 \\
0.3&64&3.753&5.784&22.1&20.8 \\
0.4&47&3.758&5.874&21.7&20.5 \\
0.5&78&3.765&5.976&18.3&17.6 \\
0.6&91&3.752&6.066&16.0&15.4 \\
0.7&141&3.755&6.132&13.4&12.9 \\
0.8&148&3.791&6.194&10.4&9.5 \\ 
\hline
\end{tabular}
\vspace{-10mm}
\end{center}
\end{table}

In this study, all of the films were grown by PLD with a KrF laser. FeSe$_{1-x}$Te$_x$ polycrystalline pellets ($x = 0 - 0.8$) were used as targets \cite{ImaiJJAP,ImaiAPEX}. The substrate temperature, laser repetition rate, and base pressure were $280\ {}^\circ\mathrm{C}$ , 20 Hz, and $10^{-7}$ Torr, respectively. 
Commercially available CaF$_2 (100)$ substrates, which are one of the most suitable materials for the thin-film growth of FeSe$_{1-x}$Te$_x$ \cite{ImaiAPEX,Hanawa,Nabeshima}, were used for the present experiments.
The films were deposited with a six-terminal shape using a metal mask for transport measurements. The measured area was 0.95 mm long and 0.2 mm wide.  We measured the temperature dependence of the electrical resistivity and Hall effect for thin films under magnetic fields of up to 9 T by using a Physical Property Measurement System (PPMS, Quantum Design, Inc.).
The specifications of the measured films are summarized in Table \ref{tab:SampleSpec}.
In this table, $x$ is the nominal composition of the polycrystalline target.
In a previous paper\cite{ImaiAPEX}, we demonstrated that the nominal Te content of the polycrystalline target was nearly identical to that of the final FeSe$_{1-x}$Te$_x$ film from the systematic change in the $c$-axis length.
The lattice parameters shown in Table \ref{tab:SampleSpec}, which were estimated from XRD measurements, are almost the same as those reported in our previous paper\cite{ImaiPNAS}.
Thus, in this paper, we use the nominal value of the Te content of the target as the film composition. 
The film thicknesses in Table I were measured using a Dektak 6-M stylus profiler.
For FeSe$_{1-x}$Te$_x$, it is known that the value of $T_\mathrm{c}$ strongly depends on the film thickness\cite{Bellingeri,Nabeshima,ImaiPNAS}.
It should be noted that the optimum film thickness for obtaining the highest $T_\mathrm{c}$ depends on the Te composition.
Therefore, we controlled the thickness of the measured film such that the highest value of $T_\mathrm{c}$ was obtained for each composition.
The values of $T_{\rm c}^{\rm onset}$ and $T_{\rm c}^{\rm zero}$ were estimated from the temperature dependence of the electrical resistivity.

\begin{figure*}[tb]
\begin{center}
\includegraphics[width=15.5cm]{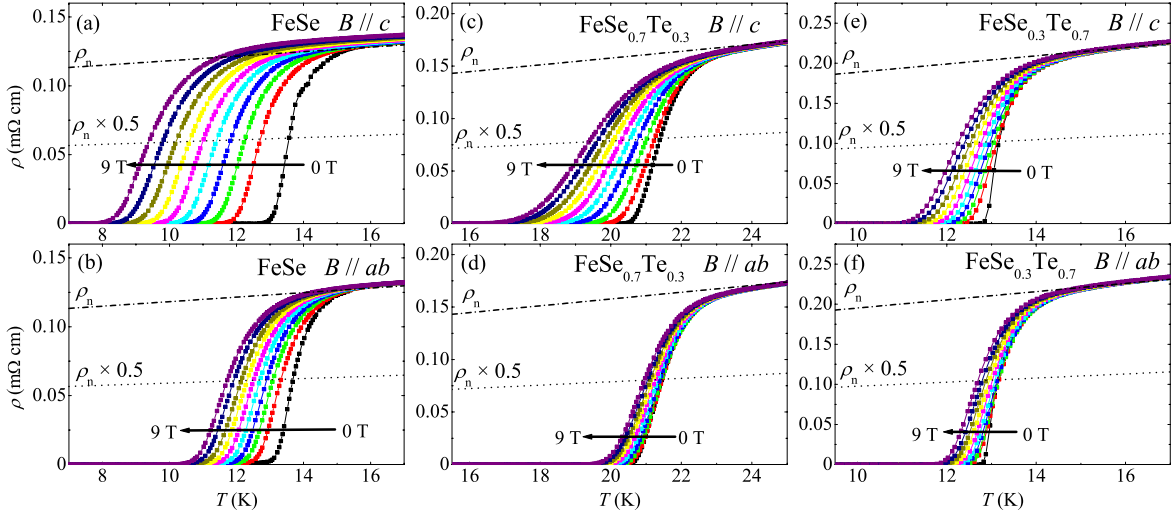}
\caption{(Color online) Temperature dependence of the electrical resistivity under magnetic fields $B$ of up to 9 T for FeSe$_{1-x}$Te$_x$ thin films with (a) $x = 0, B // c$, (b) $x = 0, B // ab$, (c) $x = 0.3, B // c$, (d) $x = 0.3, B // ab$, (e) $x = 0.7, B // c$, and (f) $x = 0.7, B // ab$.}
\label{fig:RhoTsum}
\vspace{-5mm}
\end{center}
\end{figure*}

Figure \ref{fig:RhoTsum} shows the temperature dependence of the electrical resistivity of FeSe$_{1-x}$Te$_x$ films under finite magnetic fields $B$ of up to 9 T, where the magnetic fields were applied along the $ab$ plane and $c$-axis. The results are classified into three groups (Groups A, B, and C), in terms of the Te content $x$. For Group A with $x = 0 - 0.1$, as shown in Figs.\ref{fig:RhoTsum}(a) and \ref{fig:RhoTsum}(b), where $T_{\rm c}$ is relatively low, the superconducting transition width is almost constant with increasing magnetic field (the so-called parallel shift). This parallel shift is often observed in conventional superconductors. On the other hand, for Group B with $x = 0.2 - 0.4$, as shown in Figs. \ref{fig:RhoTsum}(c) and \ref{fig:RhoTsum}(d), the superconducting transition width increases with increasing magnetic field (so-called resistive broadening), especially for $B // c$. Resistive broadening was also reported by Zhuang $et$ $al.$\cite{Zhuang} Finally, for Group C with $x = 0.5 - 1$, as shown in Figs. \ref{fig:RhoTsum}(e) and \ref{fig:RhoTsum}(f), resistive broadening is observed with increasing magnetic field. These results suggest that the nature of superconductivity is different between Group A and Groups B and C.

\begin{figure}[t]
\begin{center}
\includegraphics[width=8cm]{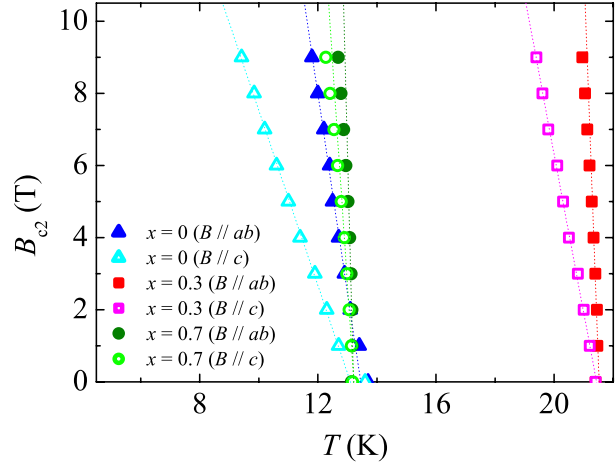}
\caption{(Color online) Temperature dependence of $B_{\rm c2}$ along the $ab$ plane and $c$-axis of FeSe$_{1-x}$Te$_x$ thin films with $x = 0, 0.3,$ and 0.7. Dotted lines are linear fits to the data.}
\label{fig:Bc2VST}
\end{center}
\end{figure}

\begin{figure}[tb]
\begin{center}
\includegraphics[width=8cm]{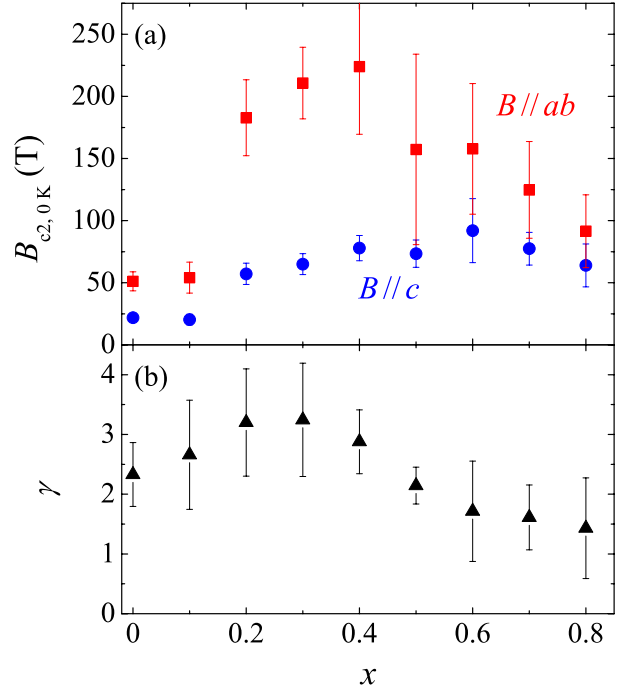}
\caption{(Color online) (a) $x$ dependence of $B_{\rm c2}$ for FeSe$_{1-x}$Te$_x$ thin films along the $ab$ plane and $c$-axis at 0 K estimated from WHH theory within the orbital limit \cite{WHH}.  (b) $x$ dependence of anisotropy $\gamma \equiv B_{\rm c2, 0 K}^{// ab}$ / $B_{\rm c2, 0 K}^{// c}$.}
\label{fig:Bc2sumforISS2}
\end{center}
\end{figure}

It is important to discuss the origins of resistive broadening for FeSe$_{1-x}$Te$_x$ films with $x \geq 0.2$. Before discussing this case, we recall the origin for high-$T_{\rm c}$ cuprate, since resistive broadening is familiar in cuprates \cite{Iye}. The origin is considered to be the result of superconducting fluctuations due to strong two-dimensionality \cite{Ikeda}. To examine whether the same scenario as cuprates is applicable for FeSe$_{1-x}$Te$_x$ thin films, we focus on the anisotropy of the upper critical field, $\gamma \equiv B_{\rm c2, 0 K}^{// ab}$ / $B_{\rm c2, 0 K}^{// c}$. Figure \ref{fig:Bc2VST} shows the temperature dependence of the upper critical field $B_{\rm c2}$ along the $ab$ plane and $c$-axis for the films with $x = 0, 0.3$, and $0.7$. For FeSe$_{1-x}$Te$_x$, the estimation of  $B_{\rm c2}$ at 0 K from low-magnetic-field data by utilizing Werthamer--Helfand--Hohenberg (WHH) theory is very difficult because this theory does not take multiband materials into account \cite{WHH}. For FeSe$_{1-x}$Te$_x$, it is widely accepted that multiple bands, which originate from Fe $3d$ orbitals, cross the Fermi level \cite{Subedi}. Moreover, the value of $B_{\rm c2}$ at low temperatures is strongly suppressed by the Pauli paramagnetic effect \cite{Kida,Lei}. 
However, in order to compare $B_{\rm c2}$ for each $x$ within the orbital limit, we consider that the orbital limit inferred using conventional WHH theory is a first-step barometer in the discussion, and we estimate $B_{\rm c2}$ at 0 K using conventional WHH theory. 
Figure \ref{fig:Bc2sumforISS2}(a) shows the $x$ dependence of $B_{\rm c2}$ at 0 K along the $ab$ plane and $c$-axis. As well as $T_{\rm c}$ for these films, the value of $B_{\rm c2}$ drastically changes between $x = 0.1$ and $0.2$. The value of $B_{\rm c2}$ for $x = 0.2$ is more than twice that for $x = 0.1$. 

Using these values, we estimate the anisotropy of the upper critical field $\gamma$. Figure \ref{fig:Bc2sumforISS2}(b) shows the $x$ dependence of $\gamma$. The value of $\gamma$ is $1.5 - 3$ and not so different between $x = 0.1$ and $0.2$, in spite of the large difference in $T_{\rm c}$ and $B_{\rm c2}$, indicating that the origin of resistive broadening for cuprates is not applicable for FeSe$_{1-x}$Te$_x$ thin films. 
In the early stages of the research on cuprates, not only superconducting fluctuations but also models of vortex motion across current lines, such as the giant-flux-creep model \cite{Tinkham}, Kosterlitz--Thouless transition model \cite{Martin}, and vortex-glass model \cite{Muller}, were proposed as possible origins of resistive broadening. 
From the above experiments, it is obvious that the nature of superconductivity changes between $x = 0.1$ and $0.2$, and further experiments are needed in order to clarify the origin of the resistive broadening \cite{Kitazawa}. 

\begin{figure}[t]
\begin{center}
\includegraphics[width=12cm]{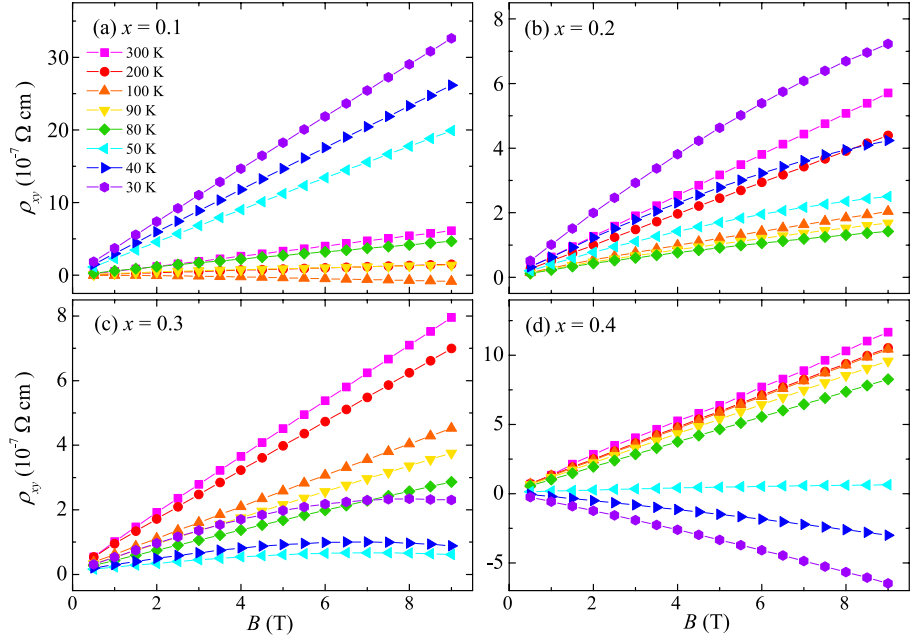}
\caption{(Color online) Magnetic field dependence of Hall resistivity for FeSe$_{1-x}$Te$_x$ thin films with (a) $x = 0.1$, (b) $x = 0.2$, (c) $x = 0.3$, and (d) $x = 0.4$.}
\label{fig:G_RhoxyH}
\vspace{-5mm}
\end{center}
\end{figure}

Next, we show the results of the Hall effect for FeSe$_{1-x}$Te$_{x}$ thin films. Figure \ref{fig:G_RhoxyH} shows the magnetic field dependence of the Hall resistivity for FeSe$_{1-x}$Te$_x$ thin films. At the lowest temperatures, the resistivity for $x = 0.2$ and $0.3$ shows nonlinear behavior as a function of the applied magnetic field. 
This behavior is the result of the multiband structure for FeSe$_{1-x}$Te$_x$ \cite{Subedi,TsukadaPRB,Hyunh}. To be precise, we should take all of the bands into account. However, we adopt a two-carrier model including one electron-type carrier (with electron density $n_{\rm e}$ and mobility $\mu_{\rm e}$) and one hole-type carrier (with hole density $n_{\rm h}$ and mobility $\mu_{\rm h}$) for simplicity. Using this model, the Hall coefficient $R_{\rm H}$, which is the slope of the Hall resistivity in the low-magnetic-field limit, can be expressed as
\begin{equation} 
R_{\rm H} = e^{-1} (n_{\rm h}  \mu_{\rm h}^2 - n_{\rm e}  \mu_{\rm e}^2) / (n_{\rm h}  \mu_{\rm h} + n_{\rm e}  \mu_{\rm e})^2.
\label{eq:rh}
\end{equation}

\begin{figure}[tb]
\begin{center}
\includegraphics[width=7cm]{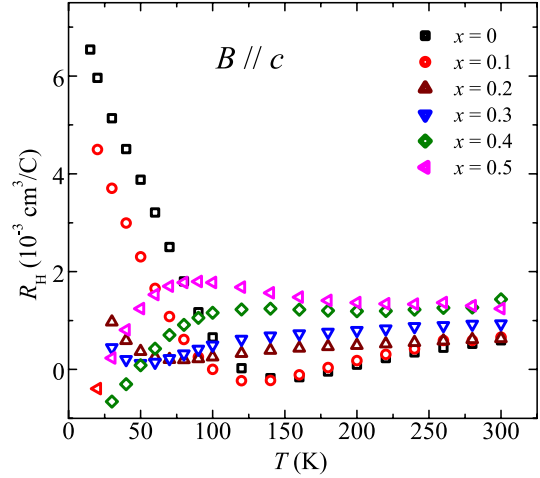}
\caption{(Color online) Temperature dependence of Hall coefficient for FeSe$_{1-x}$Te$_x$ thin films with $x = 0 - 0.5$.}
\label{fig:HallforJPSJ}
\vspace{-5mm}
\end{center}
\end{figure}

Figure \ref{fig:HallforJPSJ} shows the temperature dependence of $R_{\rm H}$ for FeSe$_{1-x}$Te$_x$ thin films with $x = 0 - 0.5$. At room temperature, the sign of $R_{\rm H}$ is positive for all films. 
Above 100 K, $R_{\rm H}$ for $x = 0$ and $0.1$ decreases as the temperature decreases, and below 100 K, it starts to increase rapidly. 
These results indicate that hole-type transport is dominant at low temperatures. 
The increase in $R_{\rm H}$ may be related to the nematicity in FeSe \cite{Horigane,Nakayama,Maeda}. In contrast, it has been reported that the sign of $R_{\rm H}$ for FeSe single crystals is negative at low temperatures \cite{Watson,Sun}. 
The origin of the difference in $R_{\rm H}$ can be explained by the difference in the band structures between single crystals and films on CaF$_2$. 
Recent angle-resolved photoemission spectroscopy measurements show that the Dirac points of FeSe single crystals are situated in the vicinity of $E_{\rm F}$. 
However, the Dirac points of FeSe films on CaF$_2$ are at some distance from $E_{\rm F}$, compared with that for FeSe single crystals \cite{Phan}. The difference in the band structures may lead to the difference in $R_{H}$ between single crystals and films on CaF$_2$ at low temperatures.

For the FeSe$_{1-x}$Te$_x$ films with $0.2 \leq x \leq 0.5$, which show large values of $T_{\rm c}$, the sudden increase in $R_{\rm H}$ is not observed below 100 K, and the values of $R_{\rm H}$ are about zero at the lowest temperature. 
From Eq. (\ref{eq:rh}), these behaviors of $R_{\rm H}$ indicate that the mobilities of the hole-type and electron-type carriers are comparable at the lowest temperatures. 
Previously, we reported the temperature dependence of $R_{\rm H}$ for films with $x = 0.5$ \cite{TsukadaPRB}, and we proposed that $T_{\rm c}$ strongly depends on the mobility of both electron-type and hole-type carriers. 
Judging from the behavior of $R_{\rm H}$ for FeSe$_{1-x}$Te$_x$ thin films with $x = 0 - 0.5$ shown in Fig. \ref{fig:HallforJPSJ}, a higher $T_{\rm c}$ is obtained when the mobilities of hole-type and electron-type carriers are comparable. This is consistent with our previous proposal\cite{TsukadaPRB}. 
The different behavior of $R_\mathrm{H}(T)$ for $x \le 0.1$ and $x \ge 0.2$ is in good agreement with the dependence of $T_\mathrm{c}$ on $x$. 
As was pointed out before, the sudden increase in $R_\mathrm{H}$ below 100 K in films with $x = 0-0.1$ is likely the result of the change in the electronic structure derived from the nematic transition. 
Thus, our results suggest that the suppression of $T_\mathrm{c}$ for $x < 0.1$ is due to the electronic nematicity. In order to further clarify the origin of the suppression of $T_\mathrm{c}$, it is important to measure the Hall resistivity under higher magnetic fields, the results of which will be discussed in a separate publication.

In conclusion, we have investigated the temperature dependence of the electrical resistivity under magnetic fields and Hall resistivity in FeSe$_{1-x}$Te$_x$ thin films. 
As well as the suppression of $T_{\rm c}$ between $x = 0.1$ and $0.2$, the superconducting transition width under a magnetic field, $B_{\rm c2}$, and the low-temperature behavior of the Hall coefficient change between $x = 0.1$ and $0.2$. Our results indicate that the sudden suppression of $T_{\rm c}$ in the range $x = 0.1 - 0.2$ is closely related to the changes in the nature of the superconductivity and electronic structure.

\begin{acknowledgments}
We would like to thank Masafumi Hanawa at CRIEPI (Central Research Institute of Electric Power Industry) for his support in the estimation of the thickness of our films and Kazunori Ueno (Department of Basic Science, the University of Tokyo) for providing the X-ray instrument. This work was supported by JSPS KAKENHI Grant Numbers 15K17697 and 26$\cdot$ 9315 and the ``Nanotechnology Platform'' (Project No.12024046) of the Ministry of Education, Culture, Sports, Science and Technology (MEXT), Japan.
\end{acknowledgments}

\end{document}